\documentclass[manuscript]{acmart}

\AtBeginDocument{%
  }

\copyrightyear{2024}
\acmYear{2024}
\setcopyright{acmlicensed}\acmConference[EuroUSEC 2024]{The 2024 European Symposium on Usable Security}{September 30-October 1, 2024}{Karlstad, Sweden}
\acmBooktitle{The 2024 European Symposium on Usable Security (EuroUSEC 2024), September 30-October 1, 2024, Karlstad, Sweden}
\acmDOI{10.1145/3688459.3688465}
\acmISBN{979-8-4007-1796-3/24/09}

\usepackage{graphicx}
\usepackage{array}
\newcolumntype{L}[1]{>{\raggedright\let\newline\\\arraybackslash\hspace{0pt}}m{#1}}
\newcolumntype{C}[1]{>{\centering\let\newline\\\arraybackslash\hspace{0pt}}m{#1}}
\newcolumntype{R}[1]{>{\raggedleft\let\newline\\\arraybackslash\hspace{0pt}}m{#1}}

\usepackage{multirow}
\usepackage{caption}
\usepackage{booktabs}
\usepackage{subcaption}
\usepackage{marginnote}

\usepackage{draftwatermark}
\SetWatermarkText{Pre-print}

\begin{document}

\title{Eyes on the Phish(er): Towards Understanding Users' Email Processing Pattern and Mental Models in Phishing Detection}



\author{Sijie Zhuo}
\email{szhu842@aucklanduni.ac.nz}
\affiliation{
  \institution{University of Auckland}
  \city{Auckland}
  \country{New Zealand}
}

\author{Robert Biddle}
\email{robert.biddle@auckland.ac.nz}
\affiliation{
  \institution{University of Auckland}
  \city{Auckland}
  \country{New Zealand}
}
\affiliation{
  \institution{Carleton University}
  \city{Ottawa}
  \country{Canada}
}

\author{Jared Daniel Recomendable}
\email{jrec291@aucklanduni.ac.nz}
\affiliation{%
  \institution{University of Auckland}
  \city{Auckland}
  \country{New Zealand}
}

\author{Giovanni Russello}
\email{g.russello@auckland.ac.nz}
\affiliation{
  \institution{University of Auckland}
  \city{Auckland}
  \country{New Zealand}
}

\author{Danielle Lottridge}
\email{d.lottridge@auckland.ac.nz}
\affiliation{
  \institution{University of Auckland}
  \city{Auckland}
  \country{New Zealand}
}

\renewcommand{\shortauthors}{Zhuo et al.}
\renewcommand{\shorttitle}{Eyes on the Phish(er)}

\begin{abstract}

Phishing emails typically masquerade themselves as reputable identities to trick people into providing sensitive information and credentials. Despite advancements in cybersecurity, attackers continuously adapt, posing ongoing threats to individuals and organisations. While email users are the last line of defence, they are not always well-prepared to detect phishing emails. This study examines how workload affects susceptibility to phishing, using eye-tracking technology to observe participants' reading patterns and interactions with tailored phishing emails. Incorporating both quantitative and qualitative analysis, we investigate users' attention to two phishing indicators, email sender and hyperlink URLs, and their reasons for assessing the trustworthiness of emails and falling for phishing emails. Our results provide concrete evidence that attention to the email sender can reduce phishing susceptibility. While we found no evidence that attention to the actual URL in the browser influences phishing detection, attention to the text masking links can increase phishing susceptibility. We also highlight how email relevance, familiarity, and visual presentation impact first impressions of email trustworthiness and phishing susceptibility.

\end{abstract}



\maketitle

\section{Introduction}

Phishing emails often mimic legitimate identities to trick users into disclosing sensitive information or clicking on malicious links. The accessibility of personal information online allows attackers to craft highly targeted phishing emails, increasing their success rate. The advancement of AI technology has further exacerbated this issue by reducing the effort required to generate sophisticated phishing emails \cite{grbic2023social, karanjai2022targeted, begou2023exploring}. 

Statistics show that about 84\% of organisations experienced at least one successful email-based phishing attack in 2022 \cite{proofprint2023}, indicating that technical solutions alone are insufficient to detect all phishing emails. Email users remain the last line of defence against these evolving phishing threats. However, users' ability to detect phishing emails is influenced by factors such as workload, stress, mood, and environmental distractions, making phishing susceptibility a dynamic and complex problem \cite{zhuo2022sok}.

These findings emphasise the importance of understanding users' email processing patterns and the mental models behind their actions. When facing phishing emails, users' split-second decisions can have catastrophic consequences.

Zhuo et al. \cite{zhuoimpact} examined the impact of workload on user interactions with phishing emails and susceptibility. They found that email relevance influences phishing susceptibility under higher workloads, but not under lower workloads. Building on this, we conducted a study using various sensors in an email processing simulation to investigate user interactions with tailored phishing emails under different workload conditions. Together with a post-study questionnaire, we studied participants' interactions with phishing emails and the mental models behind their actions.

Our study provides concrete evidence that paying attention to the email sender can reduce phishing susceptibility. We also observed that participants rarely pay attention to the hovered URL and instead check the actual phishing URL through the browser. Even then, we found no evidence that this behaviour reduces phishing susceptibility. Furthermore, we found that attention to text masking links increases phishing susceptibility. Our findings suggest that first impressions significantly impact the perceived trustworthiness of an email and the likelihood of falling for a phishing attack, offering insights into people's mental models when processing phishing emails.


\section{Related Work} \label{RelatedWork}

\subsection{Workload}

Workload has been shown to influence users' performance, with higher workload leading to increased psycho-physiological activation, strain, and fatigue, which can negatively affect task performance \cite{hockey1997compensatory, fan2017impact}. Studies indicate that both task workload and email load impact users' email reading behavior and susceptibility to phishing \cite{vishwanath2011people, jalali2020employees, rozentals2021email}. Under high workload, users find it more challenging to process emails, which can increase their susceptibility to phishing. Additionally, they are more likely to interact with phishing emails that appear relevant under high workload compared to low workload \cite{zhuoimpact}.

\subsection{Attention}

Detecting phishing emails requires attention. Users need to be somewhat suspicious of an email before considering it phishing \cite{canfield2016quantifying}, and this requires paying attention to the email, and elaborate on the content \cite{musuva2019new, harrison2016individual}. Extending this phenomenon, studying where users' allocate their attentions when processing phishing emails is valuable for uncovering the mental models behind user behaviours. 

Eye-tracking technology is one common approach to capture users' visual attention to emails to study their email processing patterns. Pfeffel et al. \cite{pfeffel2019user} used eye-tracking in an email judgement study. They found that experts are efficient: participants who were able to detect phishing spent more time looking at the header area and less time looking at emails overall. On the other hand, non-experts need more time: they took longer to look over aspects of the messages to distinguish between legitimate and phishing emails. We extend this work by studying the dwell time on specific parts of the header, such as the sender address. Further, we explored whether Pfeffel’s judgement-task findings hold within a realistic email inbox scenario.


\subsection{Phishing Email Designs}

The visual design of phishing emails plays an important role in influencing users' trust and their susceptibility to phishing attacks \cite{parsons2016users, vishwanath2011people, pfeffel2019user, burda2020testing, alseadoon2015influence}. Users tend to rely on visual cues to heuristically determine the email's legitimacy \cite{harrison2015examining}. Attackers often take advantage of this to craft phishing emails that mimics legitimate ones, to make phishing emails appear authentic, to trick users into falling for the attack \cite{williams2019persuasive}.

With the rapid growth of artificial intelligence (AI) and large language models (LLMs), many cognitively demanding tasks can be automated, improving efficiency. However, these technologies can also be exploited maliciously. Studies show that tools like ChatGPT can generate compelling email content and source code for phishing websites \cite{grbic2023social, karanjai2022targeted, roy2023generating, begou2023exploring}. Moreover, attackers can leverage AI to create sophisticated spear-phishing emails with minimal effort by providing more information about the victims to these models \cite{gradonm2023electric}.

Tailored phishing, a term introduced by Burda et al. \cite{burda2020testing}, describes phishing attacks that fall between generic phishing and spear-phishing. Similar to generic phishing, tailored phishing is a single-stage attack (hit-or-miss attack) but involves gathering additional information about the victims and their organisation to craft the phishing email, targeting a smaller population. Unlike spear-phishing, which involves iterative information gathering and attack engineering, tailored phishing requires less effort due to its single-stage nature. With advancing AI technology, we anticipate a shift from more generic, hit-or-miss attacks to more tailored and sophisticated phishing emails closely related to the recipient.



\begin{figure*}
\centering
\includegraphics[width=0.8\textwidth]{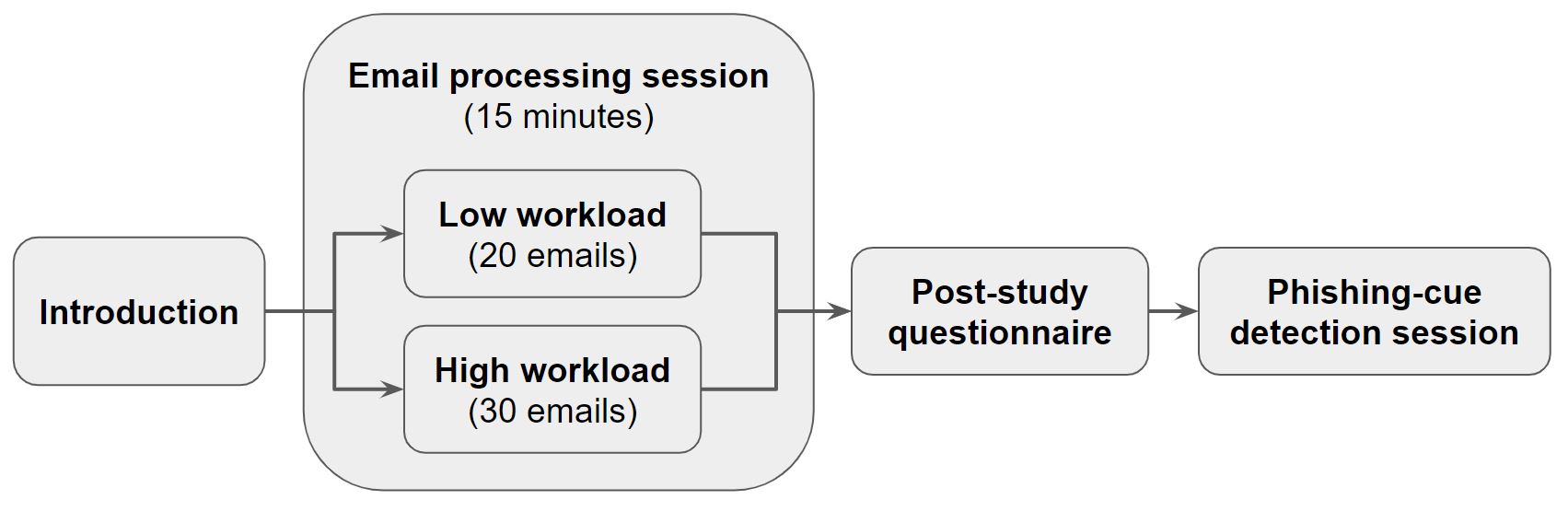}
\caption{Structure of The User Study}
 \label{fig:structure}
\end{figure*}

\section{Research Goal and Hypotheses}

Prior study indicates that users under high workload are more susceptible to relevant phishing emails compared to low workload conditions \cite{zhuoimpact}. To further explore this phenomenon, we conducted a user study aim to answer the following research question: \textbf{Why does workload influence users' susceptibility to tailored phishing emails?} We hypothesise that under low workload, users have more time to read each email (including phishing emails), making them more likely to notice phishing indicators. Conversely, a higher workload might lead users to focus on task completion rather than phishing indicators, potentially increasing their phishing susceptibility, especially when the email appears relevant to their primary task. Therefore, our first hypothesis is:


\textbf{H1: Participants under high workload are more likely to fall for tailored phishing compared to those under low workload.}
 
In this study, we distinguish between phishing indicators and phishing cues. Phishing indicators are specific, identifiable signs, while phishing cues are broader hints like tone, language, and visual presentation that raise suspicion but could also appear in legitimate emails. We track users' gaze locations when reading emails to identify areas of interest (AoI) that correspond to two phishing indicators: email sender and actual hyperlink URLs, and whether attention to these indicators affects phishing susceptibility. Email senders are key indicators because phishing emails often come from unknown senders. Recognising unfamiliar senders can help users assess the relevance or suspicion of an email, reducing phishing risk. Hyperlink URLs are critical entry points for phishing attacks, as phishing links often lead to fraudulent domains. Therefore, we propose the following hypotheses:

\textbf{H2a: People who look at the email senders are less likely to be phished.}

\textbf{H2b: People who look at the actual hyperlink URLs are less likely to be phished.}

\section{Methodology} \label{Methodology}

Our study aims to explore the impact of workload on phishing susceptibility and observe participants' interactions and mental models when dealing with tailored phishing emails. While it shares similarities with Zhuo et al.'s study, such as using the same concept of workload for phishing susceptibility, similar sensors (eye tracking and EDA), and email simulator software, our study is neither a replication nor an extension of their work. Key differences include our use of a between-subjects design instead of Zhuo’s within-subjects approach, a different set of emails and phishing stimuli, a different scenario, and a substantial post-study questionnaire. Additionally, we made the phishing emails more deceptive than those in Zhuo’s study, as they reported that most participants were not deceived under either workload condition. The structure of the user study is shown in Figure \ref{fig:structure}.

Participants were tasked with processing emails in a scenario to complete assigned primary tasks. This setup allowed us to include a secondary, implicit task -- managing other emails in the inbox, mimicking real-world complexities. This approach also enabled us to expose participants to phishing emails without explicitly priming them, avoiding bias in their susceptibility to phishing.

\subsection{Scenario and Instructions}

In the study, participants imagined themselves as temporary office workers managing emails related to a university club’s event (see Appendix for full scenario and instructions). The study targeted university students and staff, with scenario and email content designed to simulate real-life emails and activities. Participants were tasked to process emails as if they were actual recipients, handling tasks such as replying to queries and compiling event information.

Before processing the emails, participants watched an introductory video explaining the task, scenario, application interface, and other relevant background information. An information sheet with details for completing the primary task was provided, and participants were asked to familiarise themselves with it before starting the email processing session.

To enhance the realism of the scenario, we included various types of emails in the inbox, simulating a typical email inbox. We also created a website and poster for the scenario, reinforcing the idea that the information is publicly accessible.  

The true research goal was not disclosed to the participants to avoid biasing their behaviour and interaction with the emails. For the same reason, participants were not informed about the presence of phishing emails in the inbox.

\subsection{Workload Conditions}

The experiment used a between-subjects design, with participants randomly assigned to either low or high workload conditions. Both sessions lasted 15 minutes but differed in the number of emails to be processed: 20 for the low workload and 30 for the high workload. The number of task-related emails was doubled in the high workload condition to ensure participants experienced a higher workload. Our workload manipulation was similar to Zhuo's study, where they also doubled the task-related emails in the high workload condition compared to the low workload. The breakdown of the emails used in the conditions are shown in Table \ref{tab:emailSelection}.

\begin{table*}
\caption{The number of emails used in the low and high workload session}
\label{tab:emailSelection}
\centering
\begin{tabular}{|p{7cm}|>{\raggedleft\arraybackslash}p{1.8cm}|>{\raggedleft\arraybackslash}p{1.8cm}|}
\hline
\textbf{Emails}                                      & \textbf{Low workload} & \textbf{High workload} \\ \hline
Phishing emails                                      & 4& 4\\ \hline
Event-related emails                                 & 7& 14\\ \hline
Non-relevant emails (ads, internal, external emails) & 9& 12\\ \hline
Total                                                & 20                    & 30                     \\ \hline
\end{tabular}
\end{table*}

\subsection{Phishing Email Design}

We included the same four phishing emails in both low and high workload conditions to maintain consistency and comparability. The scenario simulated attackers scraping publicly accessible information and using large language models (LLMs) to generate well-crafted, tailored email messages and phishing resources, such as landing pages. We followed this process using information about the club to create tailored phishing emails for our study and finalised the designs with some manual adjustments. 

\begin{figure*}
\centering
\includegraphics[width=0.95\textwidth]{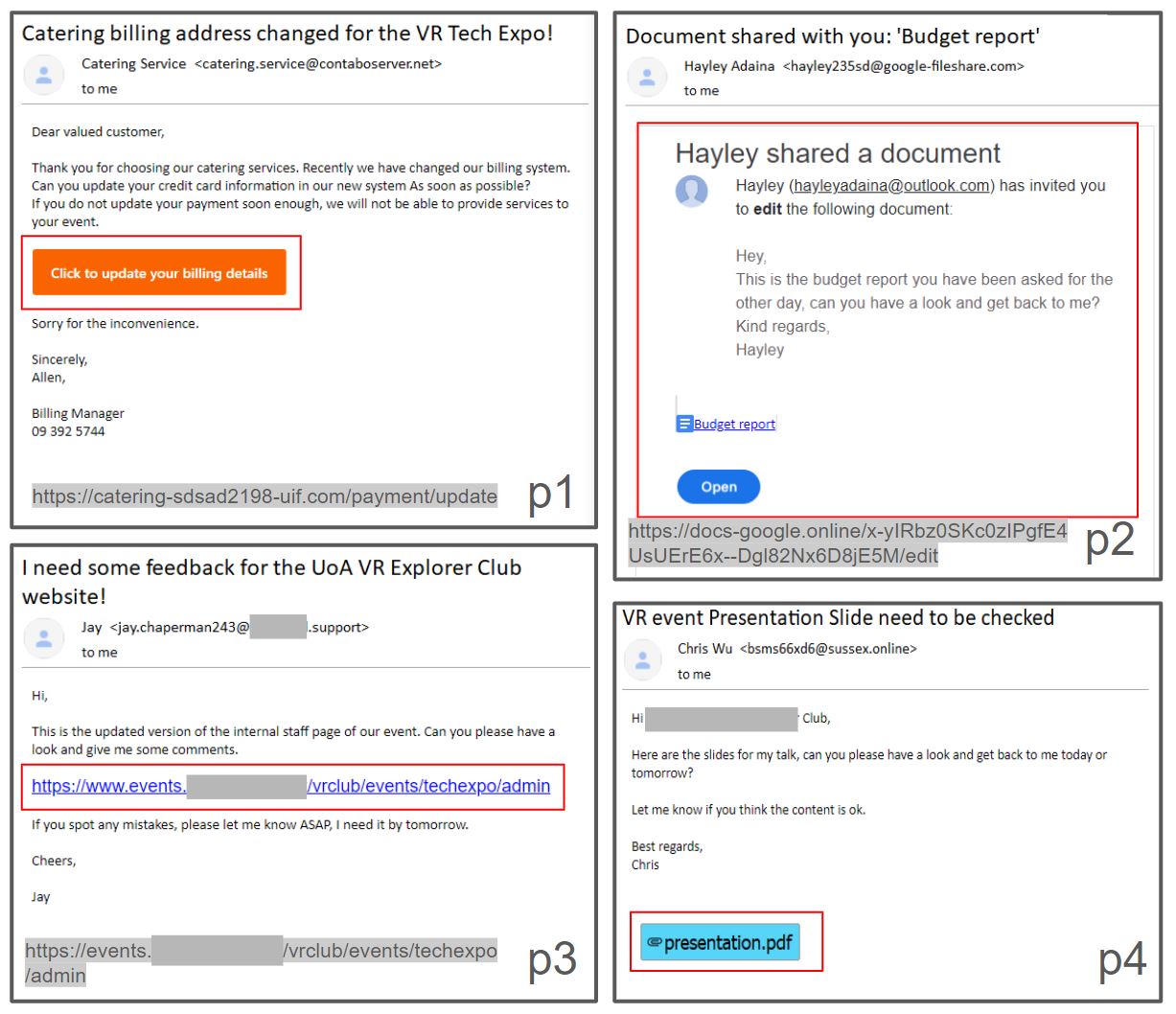}
\caption{The four phishing emails used in the study. p1) phishing email for credit card information; p2) phishing email for Google credential; p3) phishing email for university credential; p4) phishing email with attachment. The red boxes highlight the visual interests that participants tend to focus on. We removed some parts (in grey) of the email for anonymity. Note that the domain of the two links in p3 is similar and different}
 \label{fig:emails}
\end{figure*}

All phishing email senders were external unknown senders, reflecting the hit-or-miss nature of tailored attacks. Since all participants were university students or staff, they should be able to distinguish between internal and external email addresses. We ensured that all phishing emails were actionable, meaning participants could engage with them using the provided information. This design avoided scenarios like undelivered parcels, which might be easily dismissed as irrelevant to participants' tasks. We exclude spear-phishing emails and internal phishing emails because these emails would be too difficult for participants to detect in a simulated scenario. Also, they are less common because it must be tailored to each organisation attacked. The visual presentation of the phishing emails are shown in Figure \ref{fig:emails}.

In our study, participants were considered phished if they submitted their credentials to the phishing website or clicked on the phishing attachment.

\subsection{Study setup}
\begin{figure*}
\centering
\includegraphics[width=0.95\textwidth]{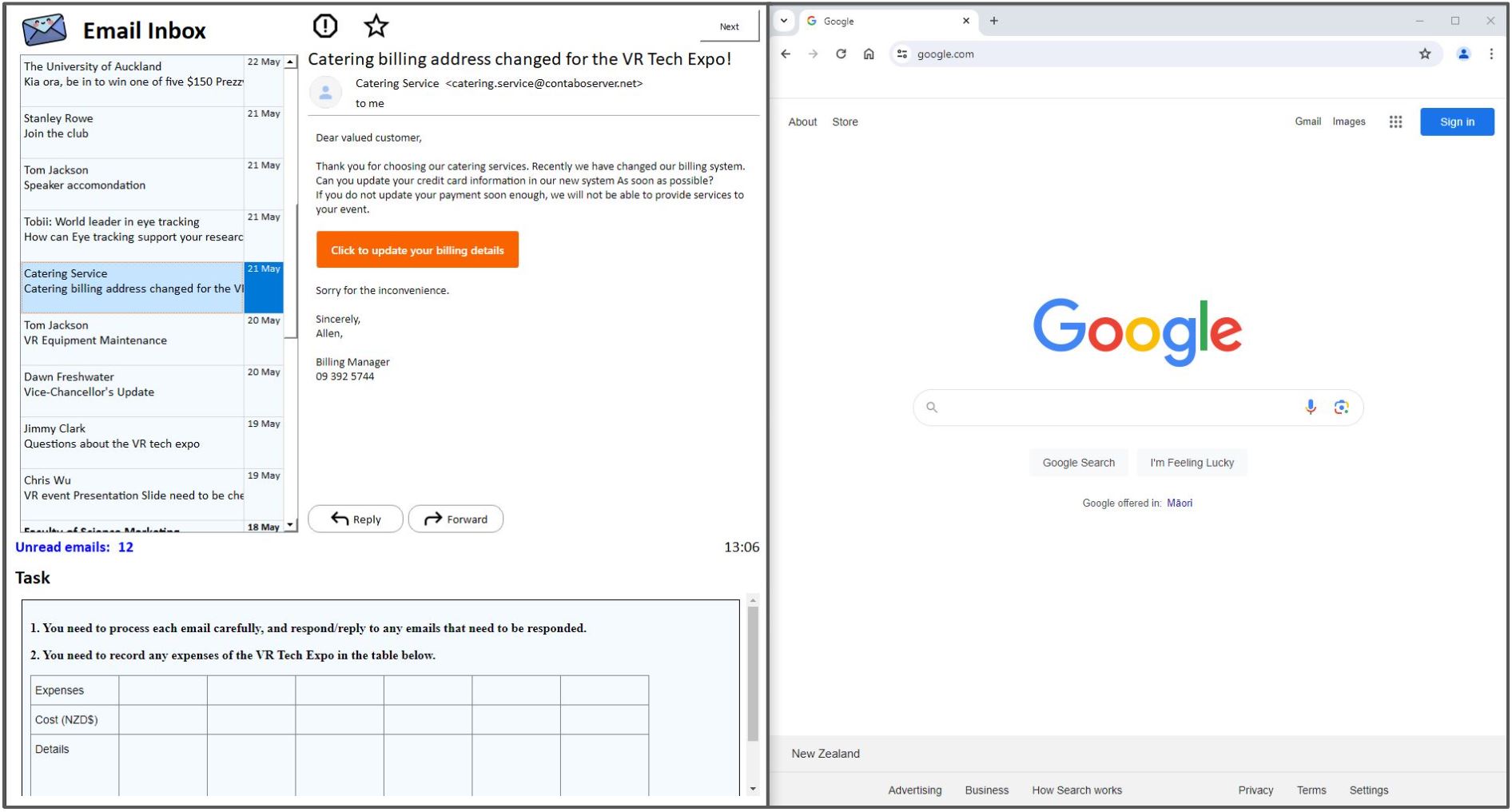}
\caption{The user interface of the email simulator. The top left section is the simulated Gmail client interface, the bottom left section is the task panel, and the right side of the screen is a browser for viewing hyperlinks and attachments.}
 \label{fig:app}
\end{figure*}

All participants completed the study in a controlled environment using the same equipment. We used the custom software from Zhuo et al. \cite{zhuoimpact}, which included a simulated Gmail client panel and a primary task panel, as shown in Figure \ref{fig:app}. The email client panel supported most interactions needed for processing emails, and participants' interactions were recorded for data analysis. The primary task panel contained content to help participants complete their primary tasks. To support accurate eye tracking, the application was positioned on the left side of the screen, allowing the right side to display embedded links and PDF attachments in a browser. Due to limited screen space, we simulated the Gmail client with the side menu bar collapsed, displaying the inbox and email content side by side.

For the hardware, we used a Tobii Pro X3-120 screen-based eye tracker to measure participants' areas of interest (AoI) and eye movement data (eye fixation rate, fixation duration) and a wrist-worn health tracker Empatica E4 to collect electrodermal activity (EDA).

\subsection{Physiological Metrics}
We used eye fixation rate, eye fixation duration and EDA to assess cognitive load. Previous studies have shown that both eye movement \cite{chen2011eye, wang2014eye, van2009uncovering} and EDA data \cite{setz2009discriminating, conway2013effect} are good indicators of cognitive load. Consistent with prior research, we measured the median fixation rate and fixation per second, and counted the average EDA peaks per minute to assess participants' cognitive load.

\subsection{Post-study Questionnaire}

We included a post-study questionnaire to understand participants' cognitive load, email reading behaviours, and interactions with various emails they had seen during the email processing session. The questionnaire consisted of five sections: (1) demographic information; (2) a NASA TLX questionnaire to assess subjective workload rating \cite{hart1988development}; (3) an email rating section; (4) a cybersecurity knowledge test; and (5) a phishing cue detection section.

The email rating section included eight emails from the email processing session, four phishing emails and four legitimate emails. For each email, participants answered four questions: (1) whether they remembered seeing the email; (2) which parts of the email they paid attention to (multiple select); (3) how they judged the trustworthiness of the email (on a 7-point scale); and (4) their justification for the email's trustworthiness. These questions helped assess participants' attention to emails and their mental models for evaluating email trustworthiness.

In the phishing cue detection section, participants received paper copies of the phishing emails from the email processing session and were debriefed on the real research goal and the existence of phishing emails in the study. They were asked to encircle any suspicious elements they noticed in the phishing emails. If they fell for a phishing email, they were asked to provide reasons for why they were deceived.

\subsection{Participants}

We recruited 115 participants, including 62 undergraduate students, 32 graduate students, and 21 staff members. The mean age was 26.2 years, with a standard deviation of 7.6. Among these participants, 60 were women, 53 were men, and 2 were non-binary. Due to some technical problems with the eye tracker, 37 participants' eye-tracking data was not collected. The 78 participants with eye-tracking data were used in the hypothesis testing.

Each participant was given a \$20 grocery voucher for their participation. Our study protocol was reviewed and approved by the Human Participants Ethics Committee of the university.

\section{Results} \label{Result}

Among the 115 participants, 57 were assigned to the low workload condition, and 58 to the high workload condition. The mean phished count for the low workload condition was 2.79 (SD = 1.00), while the mean phished count for the high workload condition was 2.60 (SD = 0.88). The breakdown of participants' phishing email performance is shown in Table \ref{tab:pEmail_breakdown}.

\begin{table*}
    \centering
\caption{Participants' phishing email performance between low and high workload condition}
\label{tab:pEmail_breakdown}
    \begin{tabular}{|c|>{\centering\arraybackslash}p{20mm}|>{\centering\arraybackslash}p{20mm}|>{\centering\arraybackslash}p{20mm}|>{\centering\arraybackslash}p{20mm}|>{\centering\arraybackslash}p{20mm}|} \hline 
         &  Did not fall for phishing&  Fall for 1 phishing email&  Fall for 2 phishing emails&  Fall for 3 phishing emails& Fall for 4 phishing emails\\ \hline 
         Low workload&  1&  5&  14&  22& 15\\ \hline 
         High workload&  0&  7&  17&  26& 8\\ \hline
    \end{tabular}
\end{table*}

We also noticed the phished rates of the four phishing emails were different. In particular, the phishing email requesting credit card information (p1) had a phished rate of 30\%, the phishing email mimicking Google Doc sharing for credential harvesting (p2) had a phished rate of 70\%, the phishing email requesting university login credentials (p3) had a phished rate of 82\%, and the phishing email pretending to be from a known sender (p4) had a phished rate of 88\%.

In the post-study questionnaire, we assessed the success of our user study scenario through three measures using 7-step scales: participants' reported familiarity with the email client interface, their perception of the realism of our scenario, and whether their experiences during the email processing session were similar to processing their own emails. A rating of 7 indicates the interface is familiar to the participant, the scenario is realistic, and the experience in the study was similar to their own experiences. We received high ratings from all three measures: $M_{familiarity} = 5.17$, $SD_{familiarity} = 1.45$, $M_{scenario} = 5.75$, $SD_{scenario} = 1.24$, $M_{experience} = 5.11$, $SD_{experience} = 1.65$. These results show our study realistically simulated a scenario that participants would experience in the real world, and they processed the emails in a way that is similar to processing their own emails.

\subsection{Workload Manipulation Check}

We evaluate the effectiveness of our workload manipulation. An alpha value of $0.05$ was used for all tests. 

To compare physiological stress in our low versus high workload conditions, we examine EDA. Mean EDA peak counts were normalised based on each participant’s mean baseline EDA peak count, which was calculated from a 15-minute session during the post-study questionnaire when participants were expected to be relaxed. We conducted two Wilcoxon signed-rank tests on the mean EDA peak count between processing emails compared to the baseline. A greater difference indicates a greater change in load. The differences between baseline and workload conditions were significant (Table \ref{tab:manipulation}), indicating both low and high workload were more physiologically arousing than baseline. We conducted a one-sided Mann-Whitney U test to directly compare EDA in low and high workload. The results were insignificant, but in the direction that is consistent with our expectations. There were more EDA peaks per minute in the high workload ($M = 2.58, SD = 2.14$) compared to low workload ($M = 1.61, SD = 1.08$), with an effect size of $r = 0.238$. 

\begin{table*}
\centering
\caption{Participants' electrodermal activity (EDA) during active email usage in high and load workload and during the baseline questionnaire}
\label{tab:manipulation}
\begin{tabular}{|p{5cm}|rr|rr|>{\raggedleft\arraybackslash}p{9mm}|>{\raggedleft\arraybackslash}p{9mm}|}
\hline
\multirow{2}{*}{EDA peaks (per min)}& \multicolumn{2}{p{2.5cm}|}{\textbf{During email interaction}} & \multicolumn{2}{p{2.5cm}|}{\textbf{During the questionnaire (baseline) session}} & \multirow{2}{*}{\textbf{W}}& \multirow{2}{*}{\textbf{p}}   \\ \cline{2-5}
             & \multicolumn{1}{r|}{Mean}   & SD     & \multicolumn{1}{r|}{Mean}   & SD      &                        &                      \\ \hline
Low workload& 4.01& 2.81& 2.94& 2.20& 159& 0.022  \\ \hline
High Workload& 3.43&2.29& 2.07& 2.03& 189& 0.005  \\ \hline
\end{tabular}
\end{table*}

We evaluated participants’ normalised median fixation duration and fixation rate from the eye tracker. Normalisation was necessary to account for individual baseline differences. Fixation duration was normalised using a min-max scaling method (value between 0 and 1). We performed one-sided, unpaired t-tests for the fixation duration, fixation rate, and the NASA TLX data. Differences were not significant. The normalised median fixation duration was in the expected direction: higher in the high workload ($M = 0.14, SD = 0.03$) and lower to low workload ($M = 0.13, SD = 0.04$), with an effect size of $r = 0.264$.

We expected that participants would spend more time reading each email under low workload compared to high workload. We performed one-sided unpaired t-tests on the average reading time of different types of emails, as shown in Table \ref{tab:readTime}. Results were significant, indicating that participants spent significantly more time reading each email under low workload compared to high workload. 

\begin{table*}
\centering
\caption{Email reading time between low and high workloads }
\label{tab:readTime}
\begin{tabular}{|p{3cm}|rr|rr|>{\raggedleft\arraybackslash}p{9mm}|>{\raggedleft\arraybackslash}p{1.2cm}|}
\hline
\multirow{2}{*}{} & \multicolumn{2}{l|}{\textbf{Low workload}} & \multicolumn{2}{l|}{\textbf{High workload}} & \multirow{2}{*}{\textbf{t(77)}}& \multirow{2}{*}{\textbf{p}}  \\ \cline{2-5}
             & \multicolumn{1}{r|}{Mean (sec)}   & SD     & \multicolumn{1}{r|}{Mean (sec)}   & SD      &                        &                     \\ \hline
Phishing emails& 27.0& 9.11& 18.9& 6.03& 4.63& < 0.001 \\ \hline
Relevant emails& 30.7&10.10& 19.8& 6.52& 5.70& < 0.001 \\ \hline
Non-relevant emails& 9.2& 6.48& 6.2& 2.68& 2.68& 0.004 \\ \hline
\textbf{All emails}& \textbf{21.4}& \textbf{5.98}& \textbf{14.4}& \textbf{4.03}& \textbf{6.04}& \textbf{< 0.001} \\ \hline
\end{tabular}
\end{table*}

Based on these results, we conclude that our workload manipulation may induce some differences in cognitive workload between low and high workload.

\subsection{H1: Phished email counts}

We hypothesised that participants under high workload are more likely to fall for tailored phishing emails compared to those under low workload. We conducted a one-sided Mann-Whitney U test to compare the number of phishing emails participants fell for between low ($M = 2.71, SD = 1.01, min = 0, max = 4$) and high ($M = 2.60, SD = 0.81, min = 1, max = 4$) workload conditions. The Mann-Whitney U test was chosen because it is suitable for comparing ordinal data between two independent groups. Our results show no significant difference, $U(N_{low} = 38, N_{high} = 40) = 834, p = 0.216$, indicating that there is no evidence that people fall for more tailored phishing emails under high workload compared to low workload. Therefore, H1 is not supported.

\subsection{H2a: Attention to email sender}

To test whether participants who looked at the email senders were less likely to fall for phishing, we examined the number of phishing emails where participants looked at the email sender (both sender name and address) and the number of times they fell for phishing. Based on the literature, it takes approximately 400 ms to recognise a word \cite{meeter2020role}. Phishing indicators such as email sender and URLs typically contain several words, so we consider 2 seconds to be a reasonable threshold to classify participants as looking at a cue. Since the data are ordinal and not normally distributed, we conducted Spearman’s correlation tests instead of t-tests. We observe a significant negative correlation, $\rho(76) = -0.362, p = 0.001$, indicating that participants who looked at the email sender were less likely to fall for phishing (Figure \ref{fig:heatmap_sender} in Appendix \ref{heatmap}; H2a supported). Of participants who did look at the sender and fall for phishing, they spent 3 seconds gazing at the sender area ($SD = 1.08$), whereas those who did not fall on average spent 3.4 seconds ($SD = 1.18$).


\subsection{H2b: Attention to actual hyperlink URLs}

We hypothesised that those who look at the actual hyperlink URLs are less likely to be phished. We evaluate this hypothesis based on three phishing emails with masked links where the actual URL can be seen through hover. The fourth contains a phishing attachment without any links. There were only two instances where participants hovered and then looked at the URLs displayed at the bottom of the screen, and in both instances, they reported the phishing email. Thus, the vast majority of participants clicked on the link without ever seeing the URL. The participant was then brought to a browser page where the URL was visible (as shown in Figure 1). Our analysis focuses on whether participants looked at the actual URL in the browser.

We performed a Spearman's correlation between the viewed URLs (through hover and in the browser) and falling for phishing (i.e., typing in credentials in the phishing website). Our result showed no significant correlation, $\rho(76) = 0.063, p = 0.585$. Thus, H2b is not supported because there is no evidence that looking at the actual hyperlink URLs influences participants' propensity to type in their credentials into a phishing site.

\subsection{Trustworthiness rating and reasoning}

In addition to the quantitative analysis we performed to study participants' attention on phishing emails, we also analysed their subjective reports on the AoIs they focused on while reading phishing and legitimate emails, and their ratings of the emails' trustworthiness and the reasoning behind these ratings.

\begin{figure*}
\centering
\includegraphics[width=0.95\textwidth]{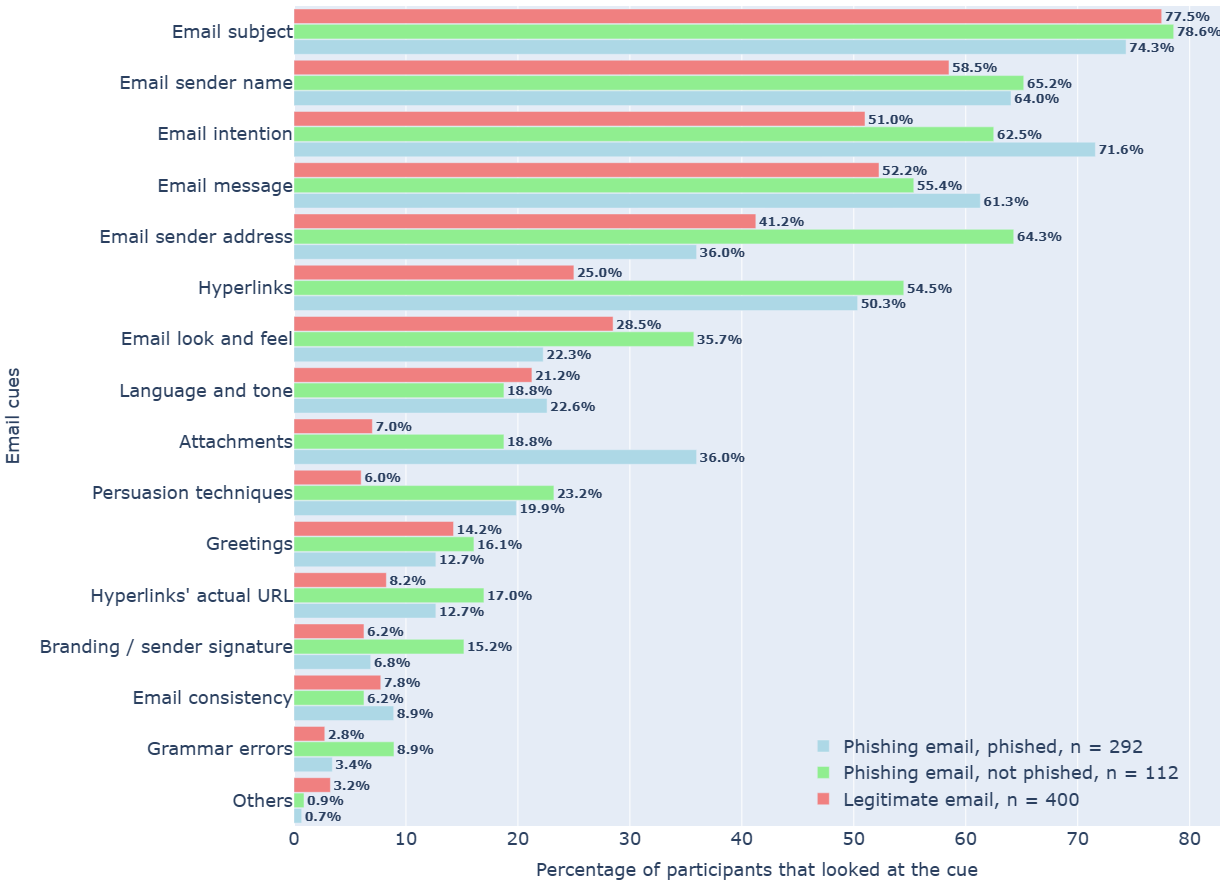}
\caption{Participants' attention on the visual elements in the phishing email}
\label{fig:cues_questionnaire}
\end{figure*}

As shown in Figure \ref{fig:cues_questionnaire}, one of the biggest differences between participants' attention to the visual elements is the attention to the email sender address. In cases where participants did not fall for phishing emails, 64\% paid attention to the email sender address, compared to only 36\% of those who fell for phishing. Additionally, although about half of the participants looked at hyperlinks, they rarely hovered over them to check the actual URLs. This is consistent with our observation that there were only two instances where participants checked the hovered URL. In comparison, when reading legitimate emails, participants focused more on the email subject and less on other visual elements.

For each of these visual elements, we conducted a Spearman's correlation test between the number of times participants reported paying attention to the visual element and the number of times they fell for phishing. The correlation was only significant between reported attention on the email sender address and the phished count, and the result is consistent with H2a, $\rho(113) = -0.260, p = 0.005$.

In the questionnaire, we asked the participants to rate the trustworthiness of the email on a scale from 1 to 7. We classify participants' ratings below 4 as trusting the email, 4 as unsure, and above 4 as not trusting the email. Participants wrote free-text responses explaining the reasons for each of the 920 ratings. In a first pass, the main author created codes as an intermediate step in a thematic analysis process \cite{braun2006using}. For example, the response ``Anything involving money and credit card is not trusted.'' is coded as ``asking for sensitive information.'' The main author then combined the codes into themes (see Appendix). The themes were tabulated in a manner consistent with verbal protocol analysis \cite{simon1984protocol}.

Based on the themes created, we performed a frequency analysis to study the most common reasons for trusting and distrusting an email. This frequency analysis helps highlight general patterns in user behaviours and perceptions and aids in understanding the users' mental models behind their decisions.

The three most mentioned themes categorised by their rated trustworthiness and actual phished stats are shown in Table \ref{tab:trustworthiness}. Participants tend to trust emails from trusted senders or when the email intention is reasonable and relevant. They would not trust emails where they can identify suspicious cues, such as emails that ask for sensitive information and are sent from unknown senders.

\begin{table*}
    \centering
    \small
\caption{Participants' reasons for rating the trustworthiness of the phishing emails}
\label{tab:trustworthiness}
    \begin{tabular}{|>{\raggedright\arraybackslash}p{5mm}|>{\centering\arraybackslash}p{2cm}|>{\centering\arraybackslash}p{2cm}|>{\centering\arraybackslash}p{2cm}|>{\centering\arraybackslash}p{2cm}|>{\raggedright\arraybackslash}p{2cm}|>{\raggedright\arraybackslash}p{2cm}|} \hline 
         \textbf{rank}&  \textbf{phished \& trusted (n = 155)}&  \textbf{phished \& not trusted (n = 14)}&  \textbf{not phished \& trusted (n = 86)}&  \textbf{not phished \& not trusted (n = 75)} & \textbf{legit \& trusted (n = 342)}&\textbf{legit \& not trusted (n = 25)}\\ \hline 
         1&  Sender trusted (73)&  Sender trusted (8)&  Sender not trusted (51)&  Sender not trusted (32)& Sender trusted (219)&Marketing email (11)\\ \hline 
         2&  Reasonable/ relevant intention (72)&  Reasonable/ relevant intention (5)&  Link/attachment did not behave as expected (21)&  Ask for sensitive information (29)& Reasonable/ relevant intention (113)&Irrelevant email (6)\\ \hline 
         3&  Links trusted (24)&  Links trusted (4)&  Inconsistent with known information (9)&  Inconsistent with known information (28)& Email looks familiar and trusted (58)&Ask for sensitive information (3)\\ \hline
    \end{tabular}
\end{table*}

It is worth noting that even though some participants noticed suspicious cues in the phishing email, they still reported trusting the email. For instance, ``sender not trusted'' was mentioned twelve times when they rated the phishing email as trusted.

Additionally, when exploring each phishing email, we found that the most common reason for not trusting p1 is \textit{asking for sensitive information} (38 times). For the other three phishing emails, \textit{suspicious email sender address} is the most frequently mentioned reason (29 times, 9 times, and 37 times, respectively).

\subsection{Reasons for falling for phishing}

After the post-study questionnaire session, participants were debriefed about the phishing emails and asked to encircle the suspicious cues in the phishing emails and provide reasons for why they fell for them. Out of the 102 participants who filled out the questions (102 * 4 = 408 entries), there were 25 phishing emails from 20 participants where they could not find any suspicious cues. This indicates that most participants know, in principle, how to identify phishing cues in phishing emails. 

Similar to the trustworthiness rating analysis, we conducted a thematic analysis by first generating 22 initial codes from participants' reasons for falling for phishing, then categorising them into eleven themes. Table \ref{tab:circle} shows the five most identified phishing cues and reasons for falling for phishing emails. The full list of themes is attached in the Appendix. Recalling H2a, we found that looking at the email sender correlates with lower phishing susceptibility. Among the 94 instances where participants fell for phishing because they trusted the phishing email sender, only 10 involved participants spending more than 2 seconds reading the sender's details.

\begin{table*}
    \centering
\caption{Participants' identified phishing cues and their reasons for falling for the phishing emails}
\label{tab:circle}
    \begin{tabular}{|>{\raggedright\arraybackslash}p{6mm}|>{\centering\arraybackslash}p{4cm}|>{\centering\arraybackslash}p{6cm}|} \hline 
         \textbf{rank}&  \textbf{Identified phishing cues (n = 408)}&  \textbf{Themes for falling for the phishing emails (n = 226)}\\ \hline 
         1&  Email sender (290)&  Sender trusted (94)\\ \hline 
         2&  Hovered URL (83)&  Time pressure (69)\\ \hline 
         3&  Text masking the link  (72)&  Reasonable/relevant intention (42)\\ \hline
 4& Persuasion techniques used in the email (51)&Trusting the visual presentation of the email/landing page (37)\\\hline
 5& URL in the browser (51)&Careless (19)\\\hline
    \end{tabular}
\end{table*}

When exploring individual phishing emails, we found that excluding the ``time pressure'' reason, participants' reasons for falling for the phishing emails differed. For p1, the most reported reason was ``reasonable/relevant intention''. For p2 and p3, it was ``trusting the visual presentation of the email/landing page''. For p4, it was ``sender trusted''.

\subsection{Post-hoc analyses}

We conducted post-hoc analyses to further explore aspects of email interactions and deepen our understanding of participants' phishing email performance. Specifically, we examined their objective and subjective reasoning for rating the trustworthiness of phishing emails and their reasons for falling for them. In this section, we present a variety of tests and again use an alpha value of 0.05. We choose not to apply corrections for multiple tests because our study involves multiple speculative tests (to assess whether participants looked, how many emails they looked at, and how long they looked at each visual element), which complicates the selection of appropriate corrections. Hence, we believe it is reasonable to report results without correction. Any results that appear significant would require further study with a priori hypotheses.



\subsubsection{Attention to clickable text masking links}

We performed a Spearman's correlation test between looking at the text masking the link and falling for phishing (i.e., entering credentials). Overall, there is significant correlation: $\rho(76) = 0.258, p = 0.005$ (Figure \ref{fig:heatmap_URL} in Appendix \ref{heatmap}). 

We used different techniques for masking the phishing links: p1 was masked with a button, p2 was masked with underlined text and a button, and p3 was masked with a legitimate URL. We performed Fisher's exact test for each phishing email to investigate the influence of these different kinds of masks. For p1 and p2, there was a positive association between looking at the mask and falling for phishing ($p_{p1} = 0.007, p_{p2} < 0.001$). There was no association for p3 ($p_{p3} = 1$). 



\subsubsection{Reading time before click}

Since participants' interactions with different phishing emails varied, we performed separate analyses for each phishing email. From the 78 participants with eye-tracking data, we selected those who clicked on the phishing link and conducted one-sided unpaired t-tests to examine the difference in email reading time before clicking between those who fell for phishing and those who did not. 

In the email mimicking a trusted template (Google Doc sharing email, p2) and the email displaying a legitimate-looking URL (p3), participants who were phished spent significantly less time reading the email before clicking (3s and 6s, respectively) compared to those who clicked but were not phished (7s and 16s, respectively). Our analysis shows no significant difference in reading time for the easy-to-detect phishing email (p1) (Table \ref{tab:readTimeB4Click}).

\begin{table*}
\centering
\caption{Participants' email reading time before clicking on the phishing link }
\label{tab:readTimeB4Click}
\begin{tabular}{|p{1.1cm}|r|r|r|r|r|r|>{\raggedleft\arraybackslash}p{10mm}|>{\raggedleft\arraybackslash}p{6mm}|}
\hline
\multirow{2}{*}{} & \multicolumn{3}{p{3cm}|}{\textbf{Clicked on the link but was not phished}} & \multicolumn{3}{p{3cm}|}{\textbf{Clicked on the link and phished}} & \multirow{2}{*}{\textbf{t}} & \multirow{2}{*}{\textbf{p}} \\ \cline{2-7}
             & n & Mean (sec)   & SD& n & Mean (sec)   & SD&                        &                    \\ \hline
p1& 14& 5.79& 5.92& 19& 8.40& 6.08& -1.23& 0.886\\ \hline
\textbf{p2}& \textbf{6}& \textbf{6.60}&\textbf{ 4.82}& \textbf{56}& \textbf{2.54}& \textbf{3.50}& \textbf{2.60}& \textbf{0.006}\\ \hline
\textbf{p3}& \textbf{3}& \textbf{15.67}& \textbf{6.60}& \textbf{64}& \textbf{5.88}& \textbf{6.59}& \textbf{2.52}& \textbf{0.007}\\ \hline

\hline
\end{tabular}
\end{table*}

We also performed tests to examine the relationship between demographic variables and phishing susceptibility. We did not find any significant results, see Appendix \ref{demographic} for more details. Similarly, we did not find a significant correlation between physiological stress and the number of times participants fall for phishing.

\section{Discussion} \label{Discussion}

Our study uncovered critical insights into user interactions with phishing emails, revealing that attention to the email sender and text masking links can significantly impact phishing susceptibility. We also found that participants primarily checked phishing URLs through the browser,  but no evidence that it would influence phishing susceptibility.

 Our findings suggest that while our workload manipulation did cause some changes in participants' cognitive load, the difference in workload conditions did not lead to significant differences in behaviour or phishing susceptibility. By examining participants' interactions with phishing emails, we reveal that people need to pay attention to phishing indicators and correctly interpret the information to detect phishing emails. However, participants' judgments can be biased by the visual presentation of the email, causing them to overlook phishing indicators.

Overall, our study provides insights into participants' mental models when processing phishing emails. We found that first impressions, influenced by email relevance, familiarity, and visual presentation, greatly impact the perceived trustworthiness of the email. This highlights the need for improved email designs and training to help users better recognize phishing attempts.

\subsection{Email sender}

Many studies highlight the importance of checking email sender addresses to assess the legitimacy of the email \cite{dixon2022holding,steves2019phish,cloudflare2023,kaspersky2021,norton2022}. Our study provides concrete evidence that looking at the email sender indeed can reduce the likelihood of falling for phishing emails. We observed many participants deliberately paying attention to the email sender address to verify the email's legitimacy. 

However, looking at the email sender address does not guarantee the detection of phishing emails. Participants also need to know how to assess the trustworthiness of the email sender. In the study, all four phishing emails utilised some form of spoofed sender to enhance trustworthiness, as shown in Figure \ref{fig:emails}. The username of p1's sender address was related to the credit card payment, p2 and p3 used fake sender domains that were similar to the real ones, and the sender name of p4 was relevant to the scenario. Our results show that participants reported 78 times across the four phishing emails that they trusted the phishing email because the sender seemed known or trusted. This suggests that while participants looked at the sender, they did not critically interpret the trustworthiness of the information. For example, in p4, many participants reported that they had seen the sender's name before, so they did not bother checking the sender address. This underlines the necessity of paying attention to details when checking emails.

\subsection{Actual hypertext URL}

Another frequently mentioned tip for detecting phishing emails is to check the hyperlinks in the email. However, our study revealed that participants rarely hovered over the hyperlinks to check the actual URL. Surprisingly, there were only two instances (from different participants) where participants checked the hovered URLs in phishing emails. In both instances, they reported the email as phishing. This suggests that those who know how to check URLs can interpret them correctly. Most participants, however, looked at the actual URL through the link in the browser. The URLs of the landing pages used spoofing techniques to enhance their trustworthiness, as shown in Figure \ref{fig:emails}. For example, the phishing URL domain for p2 was "docs-google.online," which is similar to the official Google Docs page "docs.google.com." In p3, the displayed hyperlink was a seemingly trusted URL that differed from the actual phishing URL. The high phished rates among participants who looked at the actual URL suggest they were unable to identify the phishing cue in the URL. This aligns with McAlaney and Hills's study \cite{mcalaney2020understanding}, highlighting the importance of both reading time and interpretation of visual cues in phishing detection. Additionally, the higher phished rate results from shorter reading time on the complex phishing emails (p2 and p3) before clicking also implies that impulsivity contributes to influencing participants' interaction, validating findings from Lawson et al. and Parsons et al. \cite{lawson2020email, parsons2013phishing}.

It is worth noting that viewing the URL in the browser implies they have already clicked on the link. This action is dangerous even if credentials are not entered because link clicks can be tracked by attackers and used for further malicious attempts. After opening the landing page, participants' attention could be distracted by the website content, making them less likely to check the actual URL.

\subsection{Email visual design}

The differences in phished rates between the phishing emails allow us to explore the impact of email visual designs on participants' attention and their mental models of processing phishing emails. A commonly reported reason for falling for p2 and p3 was trusting the visual presentation of the message and landing pages. After seeing familiar emails and landing pages, participants tended to assume the email was trusted and overlooked other phishing cues, such as the email sender address. This aligns with the finding that emails and websites deemed more professional-looking tended to be trusted more \cite{stojmenovic2022beauty}. This issue is becoming more serious as the cost of creating professional-looking emails and websites has decreased significantly due to advancements in AI and LLMs.

\subsection{Perception of sensitive information}

In the credit card phishing email, many participants distrusted the email because they identified the suspicious intention (asking for sensitive information) rather than focusing on the email sender address, which they typically did with other phishing emails. More than 10 participants reported not trusting the email because they simply did not trust any money-related emails. In contrast, the two phishing emails (p2 and p3) that asked for login credentials resulted in much higher phishing rates. 

Our results suggest that participants' perception of sensitive information mainly includes financial-related information, such as credit card details, while login credentials to digital services like university and Google accounts do not raise the same red flags. This distinction is troubling because it indicates a potential gap in understanding the importance of protecting non-financial information. Many participants did not realise entering credentials to phishing websites can have serious consequences. Although participants' behaviour in the experiment may not be fully reflective of real-world behaviour, the willingness to disclose credentials but not credit card details raises concerns about their perception of credentials as sensitive information.

\subsection{From phishing cues to email trustworthiness}
 
Our results show that the first 10 seconds after opening an email are crucial for participants to determine its relevance and legitimacy. For example, when the attention-grabbing element feels suspicious, such as in p1, many participants quickly raised a red flag and investigated its legitimacy. On average, participants who clicked the link but were not phished spent less time reading the email than those who got phished. This supports Pfeffel et al.'s finding \cite{pfeffel2019user} that longer email reading time does not necessarily improve phishing detection.

Conversely, when participants saw an email with a familiar Google file-sharing layout (p2), they tended to trust it without much thought. This could explain why participants spent less than 10 seconds before clicking the phishing link in p2, with those who got phished spending, on average, only \textbf{2.5} seconds on the email. This short reading time suggests that familiar layouts influenced participants' decisions. Similarly, participants could quickly identify marketing emails based on their layout and images.
 
In p4, many participants reported trusting the email because the sender looked familiar. We observed similar behaviour with legitimate emails, where participants quickly responded based on visual cues like the email sender name and layout. This suggests that the email sender name is one of the first visual elements participants focus on, significantly influencing their first impression.

These observations indicate that participants' first impressions heavily influence their judgement of an email. In the initial seconds of viewing an email, their judgements were driven by intuitions and heuristics rather than systematic evaluation. When their impression did not raise a red flag, participants tended not to critically process the email. This could explain why participants sometimes identified suspicious cues but still rated the email as trustworthy. Expanding on this idea, this may partially explain why participants did not inspect the hovered URL, as they did not question its legitimacy and it is not intuitive to view the embedded URL. Our findings reflect this, showing that without inspecting the actual URL, participants who paid attention to the text masking the link were more likely to fall for phishing.

These findings all converge on one idea, that participants' perceptions of email relevance, familiarity and visual professionalism strongly impact their first impressions of the email and its trustworthiness, which could lead to overlooking phishing indicators that could help them detect phishing emails. This reveals a critical vulnerability in how users process emails. The reliance on first impression over systematic evaluation highlights the need for further research.

Recalling that H1 is not supported, we speculate that the insignificance in the results may be due to the unsuccessful workload manipulation. We observed some significant differences, such as email reading time between conditions, but the manipulation was not strong enough to register a subjective difference between groups. Zhuo et al.'s study used a within-subject design, meaning participants directly compared their experiences of the workloads, which contributed to registering a significant subjective difference. Our study would require a much stronger workload condition to register a subjective difference. This would require adding more emails and tasks. The additional workload would increase the strength of the manipulation but also add complication to the comparison analysis. There would be a wider variation in the amount of work that people did and the number of phishing emails they would see.


\subsection{Implications}

The results of this study have several implications for future research and for improving individual and organisational defences against phishing attacks.

\subsubsection{Exploring user patterns}

Our study provides a foundation for exploring the mental models behind people's interactions with phishing emails. There is great research potential in this area. For instance, while observing where people focus their attention when reading phishing emails is important, it is also necessary to explore their action sequence, AoI sequence and the impact of these actions on their final decisions. Additionally, our study shows that merely looking at phishing indicators does not guarantee correct identification and interpretation. Future research should investigate whether users are processing the information they see or merely glancing at it, and understand why they might not process it when they should. We observed a difference in email reading time before clicking between participants who were phished and those who were not, but the reasons behind these differences remain unclear. Future studies could aim to answer these questions, providing deeper insights into user behaviour and improving phishing detection strategies.

\subsubsection{Design of security interfaces}

Our study revealed the effectiveness of paying attention to email senders in reducing phishing susceptibility and highlighted that most people do not hover over links to verify their legitimacy. This finding underscores the necessity of designing more intuitive and user-friendly email interfaces that emphasise these phishing indicators. Studies have explored various strategies to manipulate the visual presentation of the hovered URL \cite{petelka2019put, volkamer2017user}, but little has been done to highlight the email sender, especially the sender address. We believe there is great potential in improving email client interfaces to reduce users' phishing susceptibility. For instance, one could explore manipulating the visual presentation of the email sender address depending on whether the sender is known to the user.





\subsection{Limitations}

Our study has several limitations that need to be acknowledged. 

One of the main limitations is the ineffective manipulation of workload. Although we doubled the number of task-related emails under high workload conditions, the overall workload was not doubled due to the presence of other emails. The use of a between-subject design also influenced the strength of the results. We observed that many participants, experienced in processing emails, could complete the tasks in the high workload condition in under 10 minutes. Conversely, some participants struggled to complete the low workload within the 15-minute time period given.

Second, before the email processing session, participants were asked to spend several minutes reading an information sheet containing background information on the scenario and tasks. Ideally, participants should have reasonable familiarity with the content to efficiently find the needed information. However, our study did not assess participants’ familiarity with the content. As a result, some participants spent a long time reading through the document during the email processing session to find information, which influenced their interaction with the emails.

Third, some phishing emails were too relevant to the scenario, resulting in a very high phished rate. Although the phishing emails were crafted with the assumption that all the information is publicly accessible by attackers, the role-playing nature of the study blurred the line between relevant and non-relevant emails. Furthermore, since the email sender is an important cue that helps participants distinguish between legitimate and phishing emails, role-playing a character instead of processing participants' own emails meant they were uncertain whether they should know the sender.

\section{Conclusions}  \label{Conclusion}

In conclusion, this study provides valuable insights into understanding how people process phishing emails and the impact on phishing susceptibility. We demonstrated that paying attention to email sender addresses can significantly reduce the risk of being phished. However, this is not always sufficient, as users must also possess the knowledge to critically assess the trustworthiness of the sender. Our analysis of hyperlink interactions showed that participants' attention to actual URLs did not correlate to phishing susceptibility, but looking at the text masking links is correlated with a higher likelihood of falling for the phishing. Our finding suggests that people's first impression of email trustworthiness is influenced by email relevance, familiarity and visual presentation, which affects their attention to phishing indicators and, therefore, phishing susceptibility.

By integrating eye-tracking technology and realistic task simulations, we provided deeper insights into users' mental models when processing phishing emails. These findings contribute to the development of more effective cybersecurity strategies and training programs. Future research should focus on refining workload manipulation, exploring additional factors influencing phishing susceptibility, and further understanding the cognitive processes behind email interaction to build on the foundations laid by this study.


\begin{acks}

We thank the research participants for taking the time to participate in this study. We also thank the reviewers for their valuable inputs and suggestions. Robert Biddle acknowledges the support of the Natural Sciences and Engineering Research Council of Canada (NSERC), RGPIN-2022-04887.

\end{acks}

\bibliographystyle{ACM-Reference-Format}

\bibliography{references}

\break
\appendix



\section{Study scenario and instructions}

The scenario and instructions of the user study were introduced in an 8-minute introduction video; the following information was included in the information sheet to remind participants about the scenario and tasks.

Imagine you have just been given a work-from-home job related to a club called “Virtual Reality Explorer Club”. You have been asked to handle email communications and assist with gathering budget information for an upcoming event (VR Tech Expo) organised by the club. Your role involves responding to emails from club members and external parties, providing event details, and collating financial information to ensure the event's success.

You have been given the club’s email account <email address removed for anonymity reason>. \textbf{Please read and process each email carefully, ignore irrelevant emails and report any suspicious emails}.

Reply to all emails that you think need to be responded to. This would include questions related to the club, or the upcoming event, or budget information related to the upcoming event.

Collect and record the budget information related to the upcoming event in the table provided in the interface. 
After reading the scenario and instructions, the participant was invited to ask questions and clarify any aspects of the scenario or instructions that were not clear.

\section{Heatmaps} \label{heatmap}

\begin{figure}[hbt!]
\centering
\includegraphics[width=0.70\textwidth]{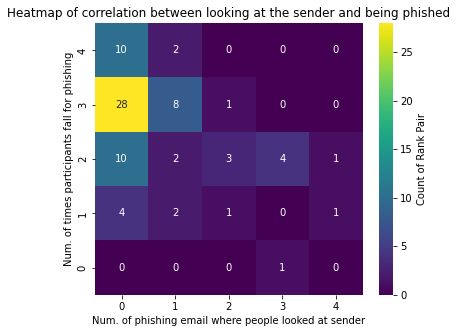}
\caption{Correlation between looking at the sender and being phished}
\label{fig:heatmap_sender}
\end{figure}

\begin{figure}[hbt!]
\centering
\includegraphics[width=0.70\textwidth]{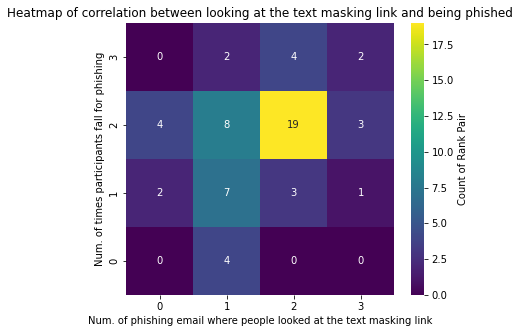}
\caption{Correlation between looking at the clickable text masking links and being phished}
\label{fig:heatmap_URL}
\end{figure}

\clearpage

\section{Demographic analysis} \label{demographic}

In the questionnaire, we collected the following demographic variables: age, gender, occupation (undergraduate, graduate student, or staff), experience with using emails, average daily email reading time, English as the first language (yes or no), tech savviness (on a scale from 1 to 7), cyber knowledge score (on a scale from 1 to 7), and the correctness of the phishing definition question.

For continuous variables (tech savvyness and cyber knowledge score), we performed a Spearman's correlation test to explore their relationship with the number of times participants fell for phishing. For categorical variables, we conducted a Kruskal-Wallis H Test. The result shows that none of these tests were significant, indicating no significant correlation between these variables and the number of times participants fall for phishing.

\section{Thematic analysis for email trustworthiness}

Based on participants' responses to four phishing emails and four legitimate emails, we created the following six themes for trusted reasons:
\begin{itemize}
    \item Sender trusted
    \item Reasonable/relevant intention
    \item Links trusted
    \item Trusted and convincing language
    \item Visual presentation of email and landing page look legitimate
    \item Email looks familiar and trusted

\end{itemize}
And we created the following six themes for not trusted reasons:
\begin{itemize}
    \item Sender not trusted
    \item Links not trusted
    \item Inconsistent with known info 
    \item Link/attachment did not behave as expected
    \item Ask for sensitive information
    \item Language used raises suspicion
    \item Unprofessional visual look and feel
\end{itemize}
We also classified three themes as being natural as the reason is not related to the trustworthiness of the email.
\begin{itemize}
    \item Marketing email
    \item Irrelevant email that would ignore
    \item     Did not consider phishing/fraud

\end{itemize}

\begin{table}[hbt!]
    \centering
    \begin{tabular}{|c|c|} \hline 
 \textbf{Themes}&\textbf{Occurrence}\\ \hline 
         Sender trusted& 73\\ \hline 
         Reasonable/relevant intention& 72\\ \hline 
         Links trusted& 24\\ \hline 
 Trusted and convincing language&17\\\hline
         Visual presentation of email and landing page look legitimate& 13\\ \hline 
         Email looks familiar and trusted& 12\\ \hline 
         Sender not trusted& 11\\ \hline 
         Link/attachment did not behave as expected& 8\\ \hline 
 Inconsistent with known info&2\\ \hline 
 Language used raises suspicion&1\\ \hline
    \end{tabular}
    \caption[Themes for Trusting Phishing Emails after Phished]{Themes for Trusting Phishing Emails after Phished, n = 155}

\end{table}

\begin{table}[hbt!]
    \centering
    \begin{tabular}{|c|c|} \hline 
 \textbf{Themes}&\textbf{Occurrence}\\ \hline 
         Sender trusted& 8\\ \hline 
         Reasonable/relevant intention& 5\\ \hline 
         Links trusted& 4\\ \hline 
         Trusted and convincing language& 4\\ \hline 
         Sender not trusted& 1\\\hline
    \end{tabular}
    \caption[Themes for Trusting Phishing Emails when not Phished]{Themes for Trusting Phishing Emails when not Phished, n =14}

\end{table}

\begin{table}[hbt!]
    \centering
    \begin{tabular}{|c|c|} \hline 
 \textbf{Themes}&\textbf{Occurrence}\\ \hline 
         Sender trusted& 57\\ \hline 
         Link/attachment did not behave as expected& 21\\ \hline 
         Inconsistent with known info& 9\\ \hline 
         Sender trusted& 8\\ \hline 
         Reasonable/relevant intention& 8\\ \hline 
         Language used raises suspicion& 8\\ \hline 
         Ask for sensitive information& 7\\ \hline 
         Links trusted& 3\\ \hline 
 Links not trusted&2\\ \hline 
 Trusted and convincing language&2\\ \hline 
 Email looks familiar and trusted&1\\ \hline 
 Unprofessional visual look and feel&1\\ \hline
    \end{tabular}
    \caption[Themes for Not Trusting Phishing Emails after Phished]{Themes for Not Trusting Phishing Emails after Phished, n= 86}

\end{table}

\begin{table}[hbt!]
    \centering
    \begin{tabular}{|c|c|} \hline 
 \textbf{Themes}&\textbf{Occurrence}\\ \hline 
         Sender not trusted& 32\\ \hline 
         Ask for sensitive information& 29\\ \hline 
         Inconsistent with known info& 28\\ \hline 
         Language used raises suspicion& 16\\ \hline 
         Links not trusted& 8\\ \hline 
         Unprofessional visual look and feel& 7\\ \hline 
         Reasonable/relevant intention& 2\\ \hline
    \end{tabular}
    \caption[Themes for Not Trusting Phishing Emails when not Phished]{Themes for Not Trusting Phishing Emails when not Phished, n = 75}

\end{table}

\begin{table}[hbt!]
    \centering
    \begin{tabular}{|c|c|} \hline 
 \textbf{Themes}&\textbf{Occurrence}\\ \hline 
         Sender trusted& 219\\ \hline 
         Reasonable/relevant intention& 113\\ \hline 
         Email looks familiar and trusted& 58\\ \hline 
         Links trusted& 35\\ \hline 
         Marketing email& 29\\ \hline 
         Irrelevant email that would ignore& 21\\ \hline 
         Trusted and convincing language& 17\\ \hline 
         Link/attachment did not behave as expected& 9\\ \hline 
 Sender not trusted&1\\ \hline 
 Unprofessional visual look and feel&1\\ \hline 
 Ask for sensitive information&1\\ \hline 
 Language used raises suspicion&1\\ \hline
    \end{tabular}
    \caption[Themes for Trusting Legitimate Emails]{Themes for Trusting Legitimate Emails, n = 342}
\end{table}

\begin{table}[hbt!]
    \centering
    \begin{tabular}{|c|c|} \hline 
 \textbf{Themes}&\textbf{Occurrence}\\ \hline 
         Marketing email& 11\\ \hline 
         Sender trusted& 4\\ \hline 
         Irrelevant email that would ignore& 4\\ \hline 
         Email looks familiar and trusted& 3\\ \hline 
         Reasonable/relevant intention& 2\\ \hline 
          Links not trusted& 2\\ \hline 
         Language used raises suspicion& 2\\ \hline 
 Sender not trusted&1\\ \hline 
 Ask for sensitive information&1\\ \hline
    \end{tabular}
    \caption[Themes for Not Trusting Legitimate Emails]{Themes for Not Trusting Legitimate Emails, n = 25}
\end{table}
\clearpage
\section{Thematic analysis for phished reasons}

\begin{table}[hbt!]
    \centering
    \begin{tabular}{|c|c|} \hline 
 \textbf{Themes}&\textbf{Occurrence}\\ \hline 
         Sender trusted& 94\\ \hline 
         Time pressure& 69\\ \hline 
         Reasonable/relevant intention& 42\\ \hline 
         Trusting the visual presentation of the email/landing page& 37\\ \hline 
         Careless& 19\\ \hline 
         Focus on the primary task& 15\\ \hline 
         Trusted and convincing language& 15\\ \hline 
         Did not consider phishing/fraud& 13\\ \hline 
 Did not check embedded links&9\\ \hline 
 Open links to know what the email is about&2\\ \hline 
 Use up information provided&1\\\hline
    \end{tabular}
    \caption[Themes for Reasons of Falling for Phishing Emails]{Themes for Reasons of Falling for Phishing Emails, n = 226}
\end{table}

\clearpage

\end{document}